# SH-SAW VOCs sensor based on ink-jet printed MWNTs / polymer nanocomposite films

H. Hallil[1], Q. Zhang[2], E. Flahaut[3], K. Pieper[1], L. Olçomendy[1], P. Coquet[2,4], C. Dejous[1], D. Rebière[1]

[1]Univ. Bordeaux, IMS, CNRS UMR 5218, Bordeaux INP / ENSEIRB-MATMECA, Talence, France

[2]CINTRA, CNRS/NTU/THALES, UMI 3288, Singapore 637553

[3]Univ. Paul Sabatier, CIRIMAT/LCMIE, CNRS UMR 5085, Toulouse, France

[4]Univ. Lille, IEMN UMR CNRS 8520 F-59652 Villeneuve d'Ascq, France

*Corresponding Authors: hamida.hallil@ims-bordeaux.fr

*Abstract*— **This study presents Shear Horizontal Surface Acoustic Wave (SH-SAW) sensor based on ink-jet printed poly (3,4-ethylenedioxythiophene) polystyrene sulfonate – multi wall carbon nanotubes (PEDOT:PSS- MWCNTs) and MWNTs-based inks as sensitive gas material. Experiments show the validation of fabrication process of acoustic platform and ink-jet printed sensitive layers. The characterization of two devices under different concentrations of ethanol vapor shows promising results in terms of reproducibility of the measurements. A sensitivity of 12H/ppm was recorded with the sensor based on ink-jet printed 600nm thickness sensitive layer.**

*Keywords—SH-SAW platform, Ink-jet printing, MWNTs, PEDOT:PSS MWCNTs, VOCs detection.*

## I. INTRODUCTION

Air pollution caused by transport and industry has become a major issue, influencing the health and social behavior all over the world. Chemical sensors for gas discrimination and quantification pose one of the outstanding challenges and need elaborate investigations on the interaction of the analyte gases and functional materials in terms of structural, chemical, and morphological attributes to better engineer the sensing properties of gas sensors [1].

Carbon nanotubes (CNTs) are a type of attractive highly sensitive material to address vapors detection due to their high structural porosity and high surface area to volume ratio [2]. CNTs walls are not reactive, but their fullerene-like tips are known to be, so end functionalization of CNTs is often used to generate functional groups (eg., -COOH, -OH, or-C=O). CNTs can be listed as three main types: Multiple Walled Carbon Nano Tubes (MWNTs), Double Walled Carbon Nano Tubes (DWNTs) and Single Walled Carbon Nano Tubes (SWNTs). These types of CNTs have different molecular geometries, which obviously imply different electrical and mechanical properties. For example, DWNTs and MWNTs have similar properties as SWNTs but they are much more resistant to chemicals. Hence, they are better suited for functionalization [3], whereas covalent functionalization of SWNTs would break some C=C double bonds, leaving holes in the structure, thus modifying electrical and mechanical properties. These carbon materials are the strongest ones yet discovered in terms of tensile strength and elastic modulus. They also have a quite low density as a solid (about 1.4 g/cm²), which leads to a high specific strength of up to 48,000 kN.m.kg-1 [4-5]. More specifically, they are used in this study for gas sensor applications, as sensitive layer to attract and immobilize different kinds of molecules like Volatile Organic Compounds (VOCs), as pollutant and industrial toxic gases.

Inkjet printing, a powerful technique for low cost electronic component fabrication, is capable of printing homogenous thin layers of nanomaterials, usually from 50nm up to a few micrometers. Particularly, one of the main difficulties is that the prepared MWNTs solution is combined into an ink that readily forms small droplets ('satellite' droplets), which are generally induced from the effects of viscoelasticity or surface tension on the drop generation of alternative MWNTs ink solutions and cause several abnormalities, roughness on the related substrate. In this paper, we present novel MWNTs ink preparation and deposition aspects, which are able to eliminate efficiently and inexpensively the difficulties named above.

This study presents a platform based on acoustic transduction associated with ink-jet printed polymer nanocomposite and MWNTs as gas sensitive layer. This material was chosen as printing ink solution, for its remarkable mechanical properties necessary to promote the propagation of the acoustic wave and for enhanced sensitivity to target species, due to high specific surface area.

## II. SH-SAW SENSOR BASED ON MWNTS / POLYMER NANOCOMPOSITE FILMS FABRICATION

### A. SH-SAW platform

The Love wave device consists of two delay lines built on an AT-cut quartz substrate. An orientation with a wave propagation direction perpendicular to the X-crystallographic axis is chosen in order to generate pure shear waves. The Love wave is generated and detected by means of interdigitated electrodes (IDTs) deposited on the substrate and prior to $SiO_2$ guiding layer, giving rise to shear horizontal surface acoustic wave (guided SH-SAW). The acoustic energy is thus confined at the near surface to maximize the sensor sensitivity. IDTs are composed of 44 split-finger pairs of gold and titanium (Ti/Au/Ti, total thickness about 150 nm) with a wavelength λ (spatial periodicity) equal to 40 μm. The geometry of the sensors plays an important role, as it affects the sensor working frequency, along with other issues. These devices are designed to operate at 117.5 MHz [6]. The sensitive layer is to be deposited on the upper delay line so as to compare the resultant response with the reference delay line (see Fig.1).

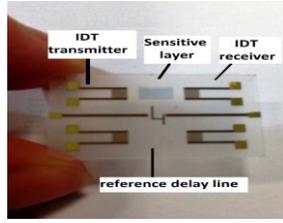

Fig. 1. Ink-jet printed coated SH-SAW device with sensitive layer of polymer nanocomposite/MWNTs

*B. Ink-jet printing and characterization of sensitive materials*

In this work we used ink-jet printing for PEDOT:PSS-MWCNTs and MWNTs-based inks deposition with a Dimatix Materials Printer (Model DMP-2800, FUJIFILM Dimatix, Inc. Santa Clara, CA). The ink homogeneity with a good dispersion of the nanotubes is a crucial issue to which a particular care has been given. Furthermore, before printing on the $SiO_2$ surface, a surface treatment is needed in order to obtain a hydrophilic surface. The surface treatment used is $O_2$-plasma, with careful control in order to prevent strong effect on the guiding layer porosity, which may impact the mechanical wave propagation, subsequently implying a lot of insertion losses as well as likely accelerated ageing due to material degradation. Besides, the effect of the surface treatment is not permanent. After investigations, the power and duration were optimized: 30W $O_2$-plasma during 30s has been defined. With these parameters, the sample becomes properly hydrophilic for about 2 hours.

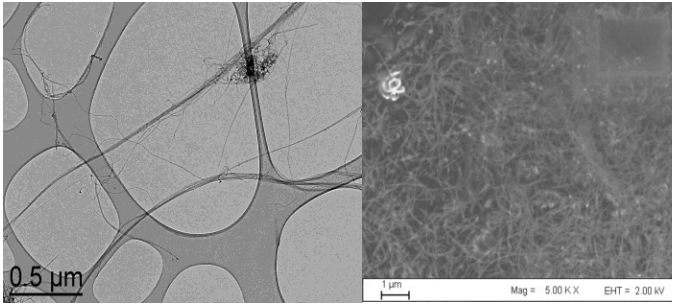

Fig. 2. (a) The TEM image of the MWNTs-based ink solution surface (b) SEM image of the printed MWCNTs films onto PEDOT:PSS-MWNTs films.

In this study, PEDOT:PSS-MWCNTs ink was chosen for commercial availability and as stable printing ink onto $SiO_2$ guiding layer surface. The commercial ink was used as a sub-layer for printing of the MWNTs-based ink solution.

MWNTs (see TEM image on Fig. 2.a) were synthesized at the CIRIMAT by Catalytic Chemical Vapour Deposition using a Co:Mo-MgO catalyst with an elemental composition of $Mg_{0.9}Co_{0.033}Mo_{0.067}O$. The catalyst was heated in an atmosphere containing 36% of $CH_4$ and 64% of $H_2$, at a total flow-rate of 15L/h, starting from room temperature to 1000°C at 5°/min. No dwell was applied and the gaseous atmosphere was maintained constant during all the procedure. The so-called nanocomposite powder obtained was processed with a concentrated aqueous solution of HCl is order to dissolve all the accessible catalyst. After thorough washings with deionised water on a filtration setup (PP membrane, 0.45μm pore size), final washings were performed with n-methylpyrrolidone (NMP) solvent. An ink was prepared in NMP with a target concentration of 400mg/L by pulsed tip-sonication for 30 minutes (1cm diameter, 30% amplitude, 5sec on; 5 sec off). The suspension was finally centrifuged to eliminate agglomerates and keep only the stable suspension in the supernatant [7].

For gas sensitive layer experiments, two layers of commercial PEDOT :PSS MWCNTs solution were ink-jet printed onto the surface of the $SiO_2$ guiding layer, followed by an additional double pass of ink-jet printed MWCNTs solution. At each one of these four coating steps, the sample's temperature during the jetting was maintained at 80°C to ensure adequate solvent evaporation. The thicknesses of the bi-layer of the polymer nanocomposite and of the bi-layer of the nanotube films were 487nm and 100nm respectively. On Fig. 2.b, From SEM images of MWNTs onto polymer nanocomposites films presented, we can clearly depict the presence of CNTs strands bonded to one another. The diameter is measured to be around 25-35nm. Such analysis further confirmed that the MWNTs have a high organization and an interesting surface morphology. Indeed, the huge surface to volume ratio will confer them outstanding sensitivity towards the molecules or particles absorbed on their surface.

*C. Electrical characterization*

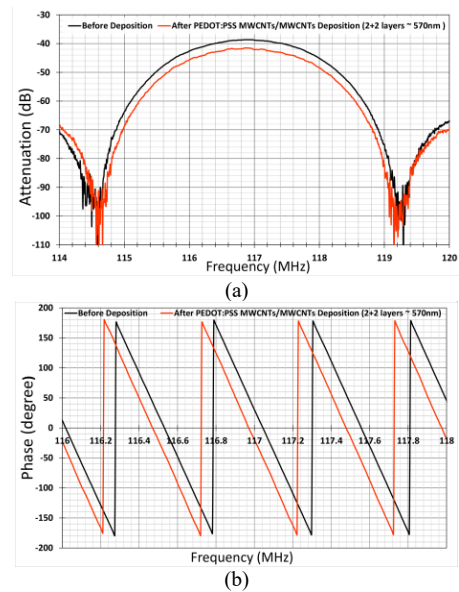

Fig. 3. Example (sample A05) (a) Gain and (b) phase, before and after inkjet-printed MWNTs/Polymer nanocomposite sensitive layers

Electrical characterization (S parameters) of the platform was performed. A wide high-accuracy spectrum has been selected on the radio-frequency network analyzer to determine the minimum of insertion loss, of both the uncoated and coated delay lines. The good performance characteristics of the ink-jet printed sensor (sample A05) are illustrated on the Fig. 3, on which are represented the transmission line characteristics (S21) in terms of the gain (figure 2.a) and the phase (figure 2.b). The sensor response was recorded before and after inkjet-printed solutions. The working frequency was observed around 117 MHz. The induced insertion losses were increased by 4dB and the equiphase frequency shift was 75.625 kHz.

## III. ETHANOL VAPOR DETECTION

The generation of ethanol vapors has been managed with a gas generator (Calibrage, PUL 010). A constant flow rate of Nitrogen, as a carrier gas (0.112 L/min) with a conventional sequence of ethanol vapors concentrations from 0 ppm to 100 ppm, circulated directly on the acoustic path of both delay lines, which were mounted into a specific cell with localized gas chamber and RF electrical connections.. In this experiment, we used two sensor samples A05 and A06 with very close characteristics in terms of working frequency. The thiknesses of the ink-jet printed layers were 570nm and 600nm for A05 and A06 samples respectively.

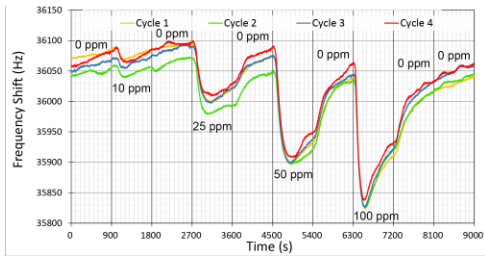

Fig. 4. Sample A05: Real-time detection of different ethanol vapors concentrations and under 4 mesurements cycles

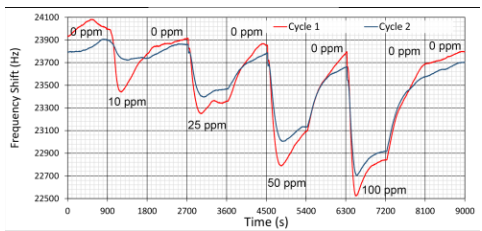

Fig. 5. Sample A06: Real-time detection of different ethanol vapors concentrations and under 2 mesurements cycles

As represented in figures 4 and 5, the real time detection of ethanol vapor for samples A05 and A06 shows clearly the evolution of the frequency shift as a function of the different concentration levels (0, 10, 0, 25, 0, 50, 0, 100 and 0 ppm) and different cycles. Thus, samples A05 and A06 exposition to different cycles show the repeatability of the measurement.

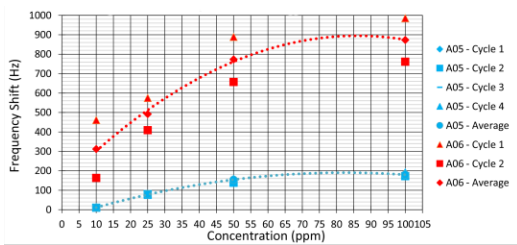

Fig. 6. Sensitivity of two samples A05 and A06 under different ethanol concetrations and cycles.

Indeed, Fig. 6, which relates the sensitivity, shows clearly these coherent responses to the different points of measurement. Also, Fig. 6 shows the experimental sensitivity values, which are estimated to be equal to 3 Hz/ppm and 12 Hz/ppm, for A05 and A06 samples under ethanol vapors, respectively. This difference could be explained by the greater thickness of the sample A06 sensitive layer compared to sample A05. These results will have to be confirmed by new ink-jet printed sensitive layers with different thicknesses of MWNTs. Finally, this study with ethanol at low concentration has put to evidence that printed MWNTs/Polymer nanocomposite exhibit sensitivities higher than alternate functional materials studied in the literature, such as $TiO_2$ (1.2 Hz/ppm) and Molecular Imprinted Polymer (MIP) (1.7 Hz/ppm) [8, 9].

## IV. CONCLUSION

SH-SAW sensor based on ink-jet printed MWNTs / polymer nanocomposite films for ethanol detection has been presented. The sensitive layer based on polymer nanocomposite and MWNTs stable inks have been printed successfully. For example, the sensitivity of the sensor sample with sensitive layer of 600nm thikness was estimated to be 12 Hz/ppm with ethanol vapors. Therefore, ink-jet printed sensitive materials on SH-SAW devices offer a real potential for application in low cost industrial processes and ambient air monitoring. As perspective, further study will focus on new deposition of sensitive films and the detection of other toxic gases regarding the specificity, along with additional ways of controlling it, such as chemical modification and signal processing.

### ACKNOWLEDGMENT

Authors are grateful to the French National Research Agency (ANR-13-BS03-0010), the French RENATECH network (French National Nanofabrication Platform), and the French Embassy of Singapore (Merlion project).